\documentclass{aastex}
\usepackage{emulateapj5,epsf}

\def\ga{\lower.5ex\hbox{$\; \buildrel > \over \sim \;$}}
\def\la{\lower.5ex\hbox{$\; \buildrel < \over \sim \;$}}
\def\lya{{\rm Ly$\alpha\;$}}
\newcommand{\msun}{{\rm M_{\odot}}}

\submitted{Accepted for publication in ApJL}
\received{2004 June 29}

\begin{document}
\shorttitle{The Nature of \lya Blobs }
\title{The Nature of \lya Blobs: Supernova-Dominated Primordial Galaxies}
\shortauthors{Mori, Umemura \& Ferrara}
\author{Masao Mori\altaffilmark{1}, 
        Masayuki Umemura\altaffilmark{2}
                   and 
        Andrea Ferrara\altaffilmark{3}}
\altaffiltext{1}{Institute of Natural Sciences, Senshu University, 
Kawasaki, Kanagawa, 214--8580, Japan.}
\altaffiltext{2}{Center for Computational Science, University of Tsukuba, 
Tsukuba, Ibaraki, 305--8577, Japan.}
\altaffiltext{3}{SISSA/International School for Advanced Studies, 
Via Beirut 4, 34014 Trieste, Italy.}
\email{mmori@isc.senshu-u.ac.jp}

%
%
%
%

\begin{abstract}
We consider a forming galaxy undergoing multitudinous 
supernova (SN) explosions, as a possible model of \lya blobs (LABs). 
For this purpose, an ultra--high resolution hydrodynamic simulation 
is performed using $1024^3$ grid points, where SN remnants
are resolved with sufficient accuracy.
It is found that multiple SN explosions produce kpc--size 
expanding hot bubbles, which drive cool, dense shells by strong shock.
The colliding high--density cooling shells radiate intensive \lya emission, 
resulting in a high \lya luminosity of $\sim10^{43}$ erg s$^{-1}$, 
comparable to the observed level in LABs.
Also, recently discovered bubbly features in some LABs are quite similar 
to the structure predicted in the present simulation.
Furthermore, the result demonstrates that LABs are 
representative of evolving primordial galaxies; they could hold direct 
information on the early chemical enrichment of galaxies, 
contrary to present--day galaxies which have undergone intense recycling 
of interstellar matter, thus erasing most of the early chemical history. 
It turns out that the metal mixing proceeds in a very inhomogeneous 
fashion, so that there appears a large spread of metallicity, 
that is,$[{\rm Fe/H}] \approx 0 {\rm ~to} -5$ or $[{\rm O/H}] \approx 1 
{\rm ~to} -4$. 
Hence, the early galactic chemical evolution may have proceeded in a different 
manner from that hitherto considered in one--zone models. 
\end{abstract}

%
%
%
%


%
%
%
%

\section{Introduction}

\lya blobs (LABs) are a new category of high redshift galaxies, which 
have been recently discovered. 
Steidel et al. (2000) found two bright \lya blobs (Blob 1 and Blob 2)
in the SSA22 field, which have
sizes of $\sim 100$ kpc and \lya luminosities of 
$\sim10^{43}$ erg s$^{-1}$ at $z=3.09$.
These LABs are associated with previously known Lyman--break galaxies 
(LBGs) at redshift $\langle z \rangle \sim 3.09$. 
Compared to typical \lya emitters found at high redshifts
(Dey et al. 1998; Weyman et al. 1998; Rhoads et al. 2000;
Hu et al. 2002; Ajiki et al. 2002; Dawson et al. 2002;
Fujita et al. 2003; Kodaira et al. 2003; Matsuda et al. 2004),
LABs are more luminous and much more extended.
LABs resemble the giant \lya nebulae associated 
with high--redshift radio galaxies, although the associated 
radio continuum flux is less than 1\%.
Recently, Matsuda et al. (2004) have found 35 candidates for
LABs in the SSA22 field. 
One third of them are apparently not associated with 
UV continuum sources that are bright enough to produce
\lya emission. Therefore, the origin of LABs is mysterious.

So far, three possibilities have been proposed as
models for LABs. The first is heavily--obscured UV sources
like an active galactic nucleus (AGN) or starburst
(Steidel et al. 2000; Chapman et al. 2004).
Chapman et al. (2004) argue that the multiwavelength observations of 
Blob 1 is at least consistent with a dust--enshrouded AGN 
surrounded by a starburst. 
The second is a cooling radiation model 
(Haiman, Spaans, \& Quataert 2000; Fardal et al. 2001). 
In a bottom--up scenario of galaxy formation, a large number 
of subgalactic halos are expected to collapse at high redshifts 
and they can emit significant \lya fluxes through collisional 
excitation of hydrogen.
Thus, the cooling radiation from proto--galaxies have been advocated 
to be a possible mechanism for observed \lya emission of LABs. 
The last is a superwind model (Taniguchi \& Shioya 2000; Ohyama et al. 2003).
Taniguchi \& Shioya (2000) have argued that 
a large--scale bipolar outflow driven by the starburst in a forming
galaxy can be responsible for the intensive \lya emission of LABs.

The deep optical spectroscopy of LABs suggests that the kinematical 
properties of LABs favor supernova (SN) driven winds (Ohyama et al. 2003). 
Interestingly, Matsuda et al. (2004) have recently revealed bubbly 
features in Blobs 1 and 2, where the bubble size is typically $\sim 15$kpc. 
The bubbly structures strongly suggest that SN events could be 
closely related to LABs. 
Therefore, complexes of various superbubbles, which can be driven by multiple 
SNe, are an attractive explanation for LABs. 
However, it has not been shown whether multiple SN explosions can
actually produce the observed LAB emission.
In this Letter, we attempt to build up a multiple SN explosion model for
LABs. 
For the purpose, we perform an ultra--high resolution ($1024^3$ grids)
simulation on the SN explosions in inhomogeneous and multiphase interstellar 
matter (ISM), where individual SN remnants are resolved. 
The results of this chemodynamical simulation,  incorporating 
spectro--photometric modeling (see 
Mori et al. 1997 and  Mori, Yoshii \& Nomoto 1999),
can be directly compared to the observations on \lya blobs.
In \S 2, the numerical model is described. 
In \S 3, we present the numerical results and 
in \S 4 the early chemical evolution of the system is discussed. 

%
%
%
%

\section{ Model}

We define a fiducial model for a \lya blob at redshift $z=3$,
in which the total mass of gaseous matter is initially assumed to be 
$M_{\rm g}=9.5\times10^{11}\,\msun$, and the total mass including dark
matter halo is $M_{\rm t}=3.7\times10^{12}\,\msun$. The baryonic 
matter to dark matter halo has been increased 
by a factor $\approx 2$ with respect to 
the cosmological value\footnote{Throughout this paper we assume a 
$\Lambda$CDM cosmology with 
$\Omega_{\rm M}=0.3$, $\Omega_{\Lambda}=0.7$, 
a Hubble constant of $h=H_0/(100$~km~s$^{-1}$~Mpc$^{-1})=0.7$ and 
$\Omega_b h^2=0.02$.}, to account for possible bias effects in rapidly 
cooling regions.  
According to the general picture of bottom--up scenarios for galaxy 
formation, we model a protogalaxy as an assemblage of numerous 
subgalactic condensations
building up the total mass of a galaxy.
As for the surface density profiles of the \lya blobs,
the galaxy as a whole is modelled as a pre--virialized system with a 
spherically--averaged density profile 
$\rho_{\rm d}(r)\propto \rho_{\rm NFW}(r)^{1/2}$ in a radius of 
106~kpc, where $\rho_{\rm NFW}(r)$ is a NFW density profile,
$\rho_{\rm NFW}(r) \propto r^{-1} (r+a)^{-2}$ 
(Navarro, Frenk \& White 1997). 
Subgalactic condensations having a zero velocity are then distributed 
to such a profile by Monte--Carlo realization.
These are taken as already virialized systems with a NFW density profile
encompassing a mass of $1.4\times10^9\,\msun$ in 
a virial radius $r_{\rm s}=8.60$ kpc. 

The star formation rate (SFR) is assumed to be proportional to the total 
gas mass. The SFR integrated over the whole system is
${\rm SFR}(t)=204 \exp (-t/t_\star ) \,\msun {\rm yr}^{-1}$ with the 
time scale $t_\star = 4.6 {\rm Gyr}$. 
Because of the short timescale of our simulation, the SFR is essentially 
constant $\sim200 ~M_\odot {\rm yr}^{-1}$ which is inferred from 
observations of bright LBGs at $z\sim3$ (Shapley et al. 2001). 
Thus, the ambiguity of $t_\star$ has little serious problem.
The mass of individual stars and the formation epoch are determined 
by randomly sampling 
both ${\rm SFR}(t)$ and
the initial mass function (IMF), 
which is assumed to be of Salpeter type with upper and lower 
mass cutoffs of $120\,\msun$ and $0.1\,\msun$, respectively. 
The newly born stars are distributed in proportion to 
the gas density using a Monte-Carlo procedure.
A star more massive than $8\,\msun$ is assumed, after the main sequence 
lifetime, to undergo a Type II SN explosion, releasing a total energy 
of $10^{51}$ erg. A mass of $2.4\msun$ of oxygen and of  
$9.05\times10^{-2}\,\msun$ of iron are ejected 
from a Type II SN explosion (Tsujimoto et al. 1995).
In the present simulation, a total of $7.53\times 10^7$ SN explosions 
occur in the whole galaxy; we focus on the first 50~Myr of the 
chemodynamical evolution of the system.

The gas dynamics is solved in three--dimensional space
with $1024^3$ Cartesian grids. Here, we employ the AUSM--DV scheme 
(Wada \& Liou 1994), which is a TVD scheme that can resolve the shock 
accurately.  The spatial resolution is 195 pc and the simulation box 
has a (physical) size of 212~kpc.  
The ejected energy and well-mixed metals by SN are supplied to 
8 local cells surrounding SN region. Then, the hydrodynamic evolution 
of metals is followed by the same algorithm as the gas density.
Since the time variation of dark matter halo potential is insignificant 
in the early stage of galaxy evolution, we assume the gravitational 
potential to be constant in time. 
Also, the gas is assumed to be optically thin 
and in collisional--ionization equilibrium. Radiative cooling is 
included self--consistently with matallicity, 
using the metallicity--dependent cooling curves by 
Sutherland \& Dopita (1993). The set of basic equations are numerically 
solved by a parallel version of the Astrophysical Fluid Dynamics (AFD2) 
scheme (see Mori, Ferrara \& Madau 2002).

%
%
%
%

\section{Results}

Figure 1 shows the numerical results at 50~Myr. 
Figures 1a and 1c show the density and temperature distributions
in a slice along the $X-Y$ plane, respectively.  After the first 
massive stars explode, the gas temperature is raised up to about $10^8$ K 
locally, and expanding hot bubbles of $\sim$kpc are produced.
Then, the strong shock drives cooled, dense shells enclosing hot bubbles.
Subsequent SN explosions further 
accelerate the expansion of hot bubbles and the ambient gas is 
continuously swept up by the dense shells; the gas density in shells 
increases further owing to the enhanced cooling. Since the density of 
ISM is low in outer regions, the bubble expansion is faster there
and SN--driven shock waves quickly collide with each other to generate 
larger bubbles of $\sim$10kpc, which are surrounded by high--density 
($\ga 0.1$ cm$^{-3}$), cool ($\sim10^4$ K) shells. 

We compute the emission properties of the galaxy assuming 
an optically thin gas in collisional ionization equilibrium and using the
MAPPINGS\,III code by Sutherland \& Dopita (1993). 
Ly$\alpha$ extinction due to dust seems negligible.
For the average gas column density, $N_H =5.6\times 10^{21}$~cm$^{-2}$, 
and metallicity in the simulation (shown below), we have 
$E(B-V)=(N_H/9.2 \times10^{21} {\rm cm}^{-2})\times 10^{[{\rm Fe/H}]}
=3.06\times 10^{-3}$. Then, using equation (20) of Pei (1992) and 
$A_B=1.2 \times 10^{-2}$, we find $\tau_{Ly\alpha}=3.1\times 10^{-2}$. 
Figures 1e and 1f show the projected distribution of the resultant \lya
emission and X--ray emission in the 0.2--2 keV band (0.8--8 keV at $z=3$),
where one arcsec corresponds to 8kpc.
It is worth noting that small bubbles with a few kpc
are smeared out in the surface brightness of \lya emission
and bubbly features of $\sim 10$kpc are prominent in outer regions. 
Such features are quite similar to the bubbly structures 
observed in Blobs 1 and 2 (Matsuda et al 2004). 
The total flux of \lya emission is 
$1.2\times10^{-16}$ erg~s$^{-1}$ cm$^{-2}$ and the corresponding
luminosity is $9.7\times10^{42}$ erg~s$^{-1}$,
which comes mainly from high--density cooling shells. This {\lya} 
luminosity is also comparable to $\sim10^{43}$ erg~s$^{-1}$ of Blobs 1 and 2.
These results suggest that LABs could be an early phase of chemical evolution
in a protogalaxy. 
The total flux of X--ray emission is 
$1.5\times10^{-19}$ erg s$^{-1}$ cm$^{-2}$ and the corresponding 
luminosity is $1.1\times10^{40}$ erg s$^{-1}$. 
This is within the constraint of the X--ray luminosity from Chandra X--ray 
observation for Blob 1, $L_X < 2.1\times10^{45}$ erg s$^{-1}$
 (Chapman et al. 2004). 

The line--of--sight velocity distribution is also shown in Figure 1d. 
Since gas in outer regions has a velocity higher than 500 km s$^{-1}$, which 
exceeds the escape velocity of this galaxy (472~km~s$^{-1}$), 
it may escape from the galaxy potential well, while in inner regions
gas velocity is lower than the escape velocity, and therefore
gas is confined within the gravitational potential.
If the total mass of galaxy is lower, winds will be more effective 
(Mori, Yoshii \& Nomoto 1999; Ferrara, Pettini \& Shchekinov 2000), and provide
an efficient mechanism to distribute heavy elements 
over cosmological volumes (Mori, Ferrara \& Madau 2002).
In the future, the more detailed bubbly structure seen in Figures 1e may be 
directly detected by high resolution observations ($\sim0.1$ arcsec),
of redshifted H$_\alpha$ emission ($1.5\times10^{41}$ erg s$^{-1}$), 
with the {\it James Webb Space Telescope}.

%
%
%
%

\section{ Discussion }

As shown in the previous section, LABs could correspond to
a quite early phase of galaxy chemical evolution.
In this case, the observed bubbly features imply that
the self--enrichment is on--going in a very inhomogeneous fashion
in LABs. 
Actually, in the present simulation, the distribution of metallicity
is very inhomogeneous as shown in Figure 1c, where
the predicted oxygen abundance [O/H] is shown.     
The gas in the vicinity of SNe is polluted with newly
synthesized heavy elements ejected from SNe, but a large amount of the 
gas still retains of low metallicity. The interactions of expanding hot
bubbles give rise to a complex structure in the inner regions, where
a metal--rich gas $[{\rm O/H}]\ga -1.74$ coexists with an almost
primordial gas. They are separated from each other by cool shells.
Interestingly, the outer regions tend to have higher metallicity compared
with inner regions. This is because the metal enriched super--bubbles are
less diluted with metal--deficient gas in the outer low--density regions.

Figure 2 shows the relation between density and metallicity, where
the abundance is represented by [Fe/H] as well as [O/H]
and the contours denote the mass level.
This figure clearly shows that the mixing of heavy elements proceeds
incompletely in an very inhomogeneous manner. 
The gas with higher metallicity of $[{\rm O/H}]> 0$
($[{\rm Fe/H}]> -0.5$) is distributed
only in low density regions of $n_{\rm g} \la 1{\rm cm}^{-3}$,
which corresponds to hot super--bubbles.
On the other hand, a part of the 
SN ejecta mix with the dense regions with primordial abundance 
(either shells or overdensities around subgalactic condensations) 
and tend to be diluted to lower metallicity of 
$[{\rm O/H}]\sim -1.7$ ($[{\rm Fe/H}]\sim -2.3$). 
Note that considerable dispersion of metallicity ($-5\la[{\rm O/H}]\la 0$) 
is found in high density regions of $n_{\rm g} \ga 1{\rm cm}^{-3}$. 
This spread of metallicity hardly depends upon the numerical 
resolution as long as hot bubbles are fully resolved in the simulation. 
Hence, the next generation of stars which are expected to form 
out of the high--density regions should exhibit
a similar dispersion of metallicity. 

So far, the theoretical models of galactic chemical evolution 
have often assumed the homogeneous ISM (one--zone model), 
with the instantaneous and perfect mixing of heavy elements 
synthesized in SNe. On the contrary, the present simulations show that
the metal mixing proceeds in a very inhomogeneous fashion.
This means that one should carefully treat the metal mixing
for the study of chemical evolution of galaxies. 

The volume--averaged metallicity of the gas in the present simulation 
is found to be $[{\rm Fe/H}]_{\rm V}=-2.30$, while the 
density--weighted metallicity is $[{\rm Fe/H}]_\rho=-2.51$. On the other 
hand, if one derives the mean metallicity based on X--ray flux, 
$[{\rm Fe/H}]_{\rm X}=-0.71$ is found, which is about forty times higher 
than the volume--averaged one. This is because X--ray is mainly emitted 
from the metal--rich gas in the hot bubbles. 
The density--weighted metallicity is well above a so--called critical 
metallicity, $Z_{\rm crit}\approx 10^{-5 \pm 1} Z_\odot$ (Omukai 2000;
Bromm et al. 2001; Schneider et al. 2002) marking the transition from a 
top--heavy IMF to a normal (i.e. Salpeter) IMF. 
Thus, the next stellar generation will form with an IMF similar to the 
present--day one, with the low--mass stars possibly forming the 
metal--poor halo population 
(McWilliam et al. 1995; Ryan, Norris \& Beers 1996). In addition, it is 
worth noting the existence of very low metallicity 
($[{\rm Fe/H}] \approx -5$), high density ($n_{\rm g}\sim 1$ cm$^{-3}$) 
regions. Since the abundance of these regions can be 
below the critical metallicity, 
the IMF could be similar to that of Pop III stars 
(e.g. Nakamura \& Umemura 2001). 
Recently, an extremely metal--poor star 
($[{\rm Fe/H}] \approx -5.3$) is discovered by Christlieb et al. (2002) 
may be a relic of such stars. 
It should be kept in mind, however, that the model galaxy 
is constructed from primordial gas in this paper.
At redshift $z=3$, \lya absorption lines in quasar spectra show that
at least some fraction of the IGM is polluted to $10^{-2.5} Z_\odot$. 
If a galaxy forms from a well--mixed pre--enriched 
medium, the metallicity would be accordingly shifted to higher values, 
hence precluding the formation of extremely metal--poor stars.

\section{Conclusions}

We have suggested that LABs can be identified with primordial galaxies
catched in a supernova-dominated phase. This conclusion is based on
ultra--high resolution hydrodynamic calculations which resolve individual
SN remnants.
As a result, we found that the resultant \lya luminosity
can account for the observed luminosity of LABs; in addition,
the bubbly structures produced by multiple SN explosions are
quite similar to the observed features in \lya surface brightness
of LABs. Hence the emerging theoretical picture is fully consistent
with experimental data.
If this picture is correct, bubbly features may show 
that the metal mixing is highly inhomogeneous
in LABs. The present simulation predicts a quite wide range
of metallicity for the next generation of stars. This may
affect the galaxy evolution significantly.

%
%
%
%

\acknowledgments
\noindent This work was supported in part by the Grant--in--Aid of the 
Ministry of Education, Culture, Science, and Sports, 14740132 and 16002003,
and by the Promotion and Mutual Aid Corporation for Private Schools of Japan.
The numerical computations were carried on a massive parallel computer 
CP--PACS at the CCP, University of Tsukuba, using 1024 processors, and
the data analysis was done by a parallel computer SPACE at Senshu university.

%
%
%
%

%
%
%
%

\clearpage
\begin{figure}
\epsscale{0.8}
\plotone{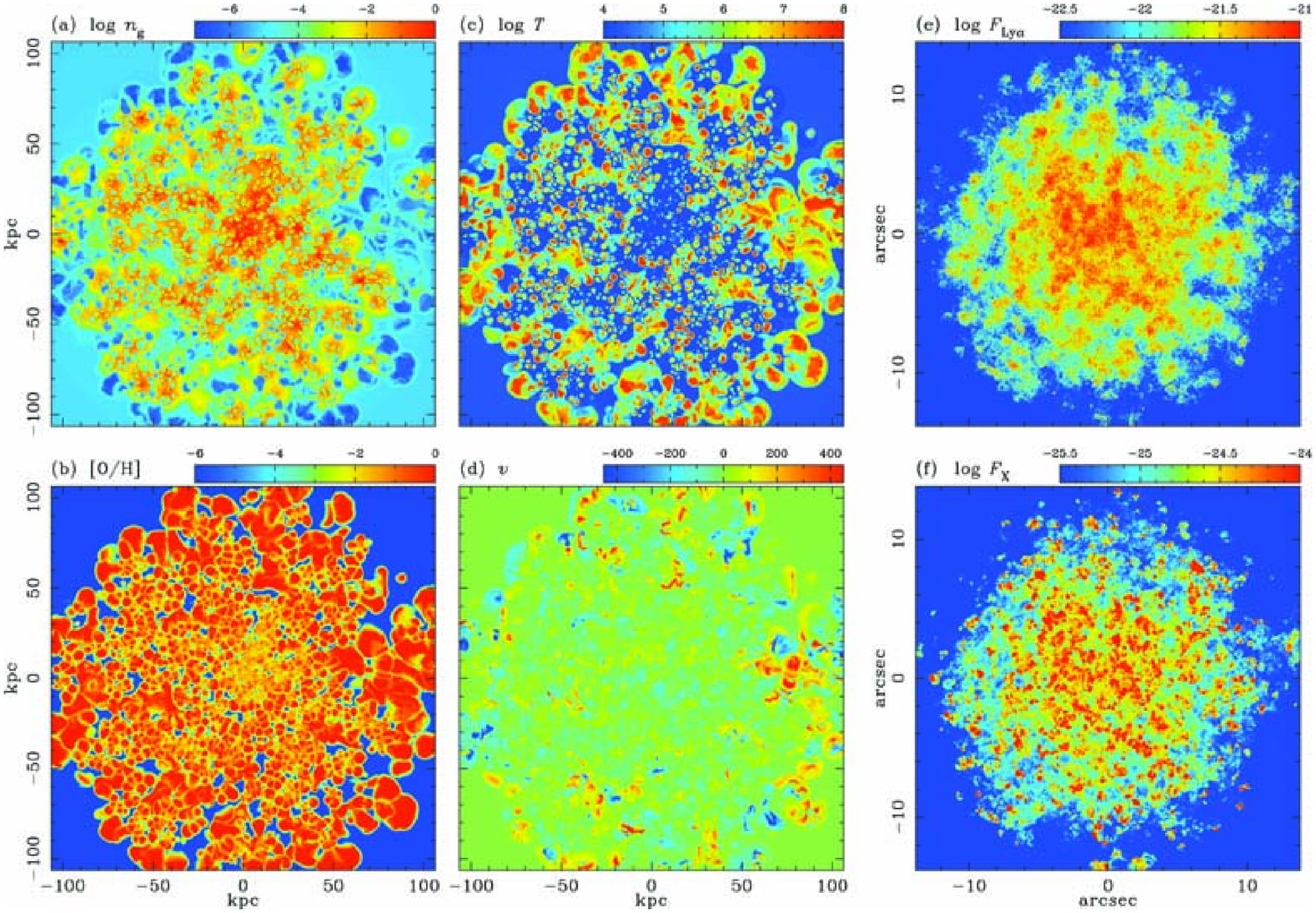}
\epsscale{1.0}
\figcaption{Numerical results at 50 Myrs on the multiple SN explosions
in a forming galaxy. Here are shown the sectional distributions of
(a) the density, (b) the oxygen abundance [O/H], 
(c) the temperature, and (d) the line--of--sight velocity 
in a slice along the {\it X--Y} plane. 
Also, (e) the projected distribution of the \lya 
emission and (f) that of the soft X--ray emission in the 0.2--2 keV band
(0.8--8 keV at the redshift z=3) are shown. Here
the angular resolution is $2.7\times10^{-2}$ arcsec (corresponding 
to the numerical resolution, 195 pc).
\label{fig1}}
\end{figure}

\clearpage
\begin{figure}
\plotone{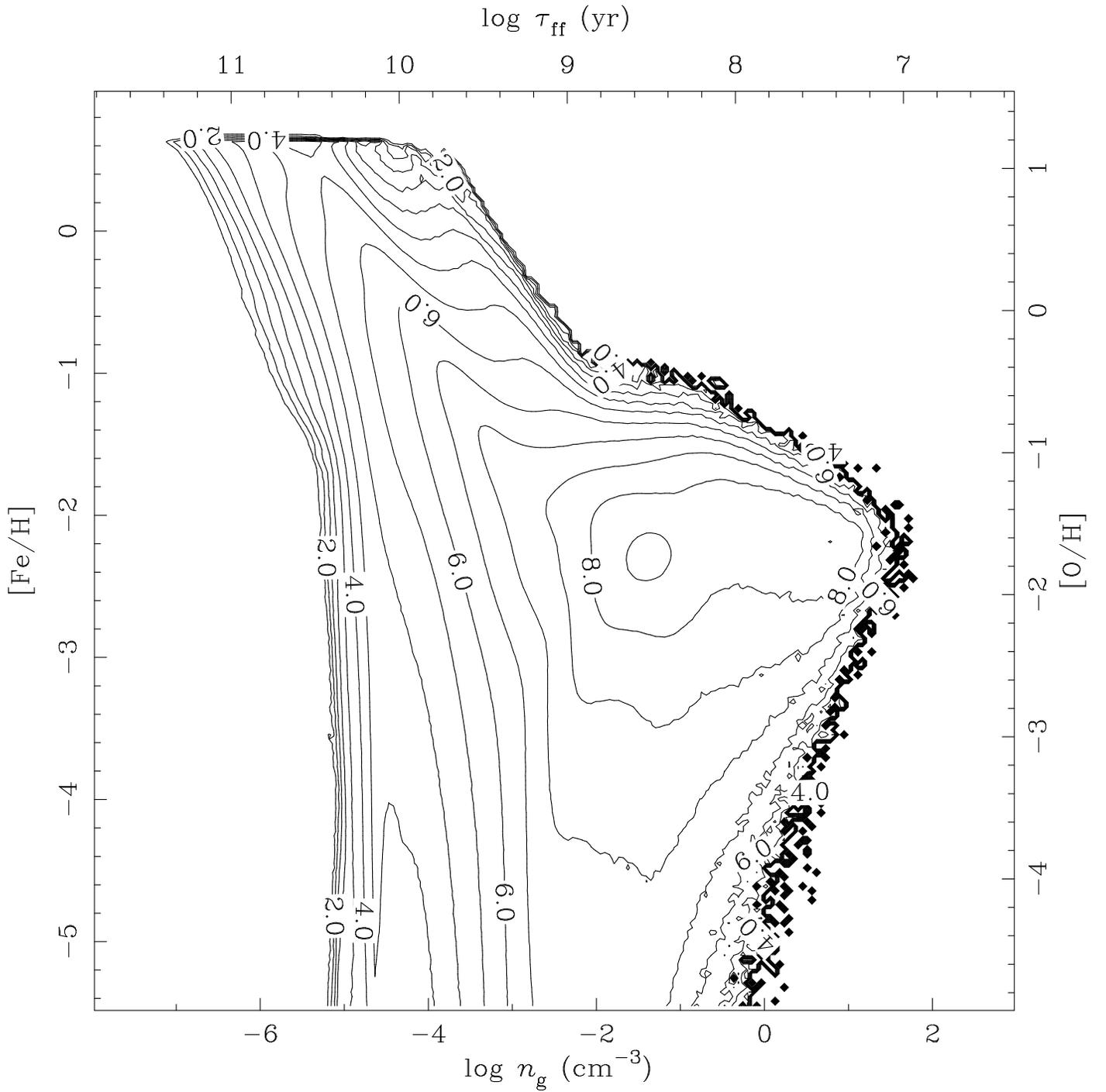}
\figcaption{Relation between gas density and metallicity.
The contours show mass levels in logarithmic scales in units of $\msun$.
Oxygen abundance [O/H] corresponds to iron abundance of [Fe/H] =[O/H] -0.559 
for the yield adopted.
\label{fig2}}
\end{figure}

%

\end{document}